\documentclass[onecolumn,showpacs,preprintnumbers,amsmath,amssymb]{revtex4}
 \usepackage{graphicx}
\usepackage{latexsym} 
\usepackage{dcolumn}
\usepackage{graphicx, subfigure}

\usepackage{graphicx}
\usepackage{dcolumn}
\usepackage{bm}

\usepackage{bm}

\begin{document}

\newcommand{\eq}{\begin{equation}}                                                                         
\newcommand{\eqe}{\end{equation}}              

\title{Heat conduction: hyperbolic self-similar shock-waves in solids} 

\author{ I. F. Barna$^a$ and R. Kersner$^b$}
\address{$^a$  Energy Research Centre of the Hungarian Academy 
of Sciences, \\ (KFKI-AEKI), H-1525 Budapest, P.O. Box 49, Hungary, \\ 
 $^b$University of P\'ecs, PMMK, Department of Mathematics and Informatics, 
Boszork\'any u. 2, P\'ecs, Hungary}

\date{\today}

\begin{abstract} 
Analytic solutions for cylindrical thermal waves in solid medium is given based on the nonlinear hyperbolic system 
of heat flux relaxation and energy conservation equations. The Fourier-Cattaneo phenomenological law 
is generalized where the relaxation time and heat propagation coefficient have a general power law temperature dependence.
From such laws one cannot form a second order parabolic or telegraph-type equation. 
We consider the original non-linear hyperbolic system itself with the self-similar Ansatz for the temperature distribution 
and for the heat flux. As  results continuous and shock-wave solutions are presented. For physical establishment numerous materials with various temperature dependent heat conduction coefficients are mentioned.  

\end{abstract}

\pacs{44.90.+c, 02.30.Jr}
\maketitle
In contemporary heat transport theory (ever since Maxwell's paper \cite{maxw}) it is widely accepted in the literature that 
only for stationary and weakly non-stationary temperature fields the constitutive equation assumes that a temperature 
gradient $\nabla T$ instantaneously produces heat flux ${\bf{q}}$  according to the Fourier law 
\eq
{\bf{q}}({\bf{x}},t) = -\kappa {\bf{ \nabla }}T({\bf{x}},t).
\label{fur}
\eqe
Combining this equation with the energy conservation law the usual parabolic heat conduction equation is given. 
Heat conduction mechanisms can be classified via the temperature dependence of the coefficient $\kappa \sim T^{\nu} $.
There are three different cases of thermal conductivity,  normal heat conduction which obeys the Fourier law $(\nu =0)$, 
slow $(\nu > 0)$  and fast heat conduction $ -2 <  \nu < 0$. 

In plasma physics if the temperature range is between $10^5$K and  $10^8$ K then the 
coefficient of the heat conductivity $\kappa$ depends on the temperature 
and density of the material. It is usually assumed to have a power dependence
$ \kappa =  \kappa_0 T^{\nu}v^{\mu} $
where $v=1/\rho$ is the specific volume the coefficient $\kappa_0$ and 
the exponents $\nu, \mu$ depend on the heat conduction mechanism \cite{zeld}.  
With radiation heat conduction one has $  4 \le \nu \le 6, \hspace*{2mm}  
1 \le \mu \le 2$; with electron heat conduction and fully ionized plasma 
$\nu = 5/2, \hspace*{2mm}\mu=0$. 
For magnetically confined non-neutral plasma the classical heat conduction 
coefficient is the following \cite{dubin} 
$\kappa \approx \frac{c_1}{\sqrt{T}}ln[c_2 T^{3/2}] $. 
Parabolic thermal wave theory is based on this approach \cite{zeld,zk}.
In plasmas heat conduction is strongly coupled to flow properties which we will not consider in the 
following. The linear parabolic theory predicts infinite speed of propagation which is known as the "paradox of heat conduction" (PHC).
The following two theories resolve this contradiction. 
 
However, if the time scale of local temperature variation is very small, 
Eq. (\ref{fur}) is replaced by  
\eq
{\bf{q}}({\bf{x}},t+\tau) = -\kappa \nabla T({\bf{x}},t) 
\eqe
where $\tau$ is called the thermal relaxation time. This is a thermodynamic property of the materials 
which was determined experimentally for large number of materials. Although $\tau$ turns out to be 
very small in many instances e.g. is of order of picoseconds 
for most metals, there are several materials where this is not the case, most notably sand (21 s), H acid (25 s), NaHCO$_3$ (29 s), and biological tissue (1-100 s)  \cite{ind}.

Unlike the Fourier's heat conduction law, this constitutive equation is non-local in time. 
The desired local character can be restored with the Taylor expansion of ${\bf{q}}$ by time which is usually truncated 
at the first order namely
\eq   
{\bf{ q}}({\bf{x}},t) + \tau \frac{\partial {\bf{q}}({\bf{x}},t) }{\partial t} = -\kappa \nabla  T({\bf{x}},t). 
\eqe
This is the well-known Cattaneo heat conduction law \cite{cat} the second term on the left hand side is known as the
"thermal intertia". (Unfortunately, this form is not Galilean invariant, and gives a paradoxial results if the media is in motion,
this problem was eliminated in by \cite{christov}.)
Combining this constitutive equation with the energy conservation yields the hyperbolic telegraph heat conduction equation 
where $\tau$ and $\kappa$ are constants.
Hyperbolic equations usually ensure finite propagation velocity. 
Unfortunately, telegraph equations has no self-similar solutions which would be a desirable physical property. 
In the work of \cite{barn} a non-autonomous telegraph-type heat conduction equation is presented with 
self-similar non-oscillating compactly supported solutions. 
A review with a large number of physical models of heat waves can be found in \cite{ind, prec}. 
A recent work  on the speed of heat waves was published by \cite{makai}. 

Our starting point is the following

\begin{eqnarray}
q_t  &=& -\frac{q}{\tau} - \frac{\kappa}{\tau} T_r,  
\label{eqn000} \\
c_0 T_t &=& -q_r 
- \frac{q}{r}.   \label{eqn00} 
\end{eqnarray}

The first equation of the system is the generalized  Fourier-Cattaneo 
heat conduction law and the second one is the energy conservation condition 
for the radial coordinate.  The heat flux $q = q(r,t)$ and the temperature dependence $T = T(r,t)$ have 
radial coordinate and time dependence.  The subscripts r and t notate the partial derivatives with respect to the radial 
coordinate and the time, respectively. (From now on we investigate the radial coordinate of a cylindrical symmetric problem 
as spatial dependence.) 
The parameter  $c_0 = \rho c$ where $\rho$ is the mass density and $c$ is the 
specific heat. Second order effects such as compressibility are neglected ($\rho$ and are c constants during the process). 

In the following we shall suppose that the heat conduction coefficient and the thermal relaxation depend on temperature
on the following way 
\eq
\kappa = \kappa_0T^{\omega},   \hspace*{1cm} \tau = \tau_0 T^{-\epsilon}.
\eqe

The $\kappa_0$ and $\tau_0$ are real numbers with the proper physical dimensions. 
Now our dimensionless system reads 

\begin{eqnarray}
q_t&=& - T^{\epsilon}q -  T^{\epsilon +\omega} T_r,  
\label{eqn0} \\ 
T_t &=& - q_r - \frac{q}{r}. \label{eqn1}   
\end{eqnarray}
There are various phenomenological heat conduction laws available for all kind of solids,  
without completeness we mention some well-known examples. 
For pure metals according to \cite{jones} (Page 275 Eq. 27.3) the Wiedemann-Franz low  the thermal conductivity is proportional with the 
electrical conductivity $\sigma$ times the temperature 
$\kappa = \sigma L T.  $ The proportionality constant L is the so called Lorentz number with the approximate
numerical value of $ 2.44 \times 10^{-8} W\Omega K^{-2} $.  For exact numerical data for various metals see 
\cite{ashroft}. The relaxation time $\tau$ is proportional to the heat conduction coefficient divided by the temperature. 
For metals with impurities the thermal resistivity (inverse of the thermal conductivity) is $
\kappa^{-1} = AT^2 + BT^{-1} $ 
where A and B can be obtained from microscopic calculation based on quantum mechanics  \cite{jones}  (Page 297 Eq. 40.11). 

A hard-sphere model for dense fluids from \cite{netl} derives a relation where the heat flux $q(x,t) = a\nabla T(x,t) + q^2(x,t)$ which certainly means a non-linear heat propagation process. 
For the heat conduction in nanofluid suspensions \cite{vadasz} derives the $\kappa \approx  c/(T_2-T_1)$ law with additional time dependence.  Another exotic and very promising new materials are the  carbon nanotubes 
which have exotic heat conduction properties.  Small {\it{et al.}} \cite{small}  performed heat conductivity measurements 
and found that at low temperatures there are two distinct regimes 
  $\kappa(T)  \sim T^{2.5}  \>\>  (T < 50 K)$ and 
$\kappa(t)  \sim T^2 \>\>  (50 < T <150 K) $. Beyond this regime there is a deviance from this quadratic 
temperature dependence and the maximum $\kappa$ value lies at 320 K. 
Above this value - at large temperatures -  there is a $\kappa(T)  \sim 1/T$ dependence according to 
\cite{berber}.
 Additional nanoscale systems (like silicon films, or multiwall carbon nanotubes) have exotic temperature 
dependent heat conduction coefficients as well, for more see \cite{cahill}. 
For encased graphene the heat conduction coefficient is $\kappa \sim T^{\beta}$ where $1.5 <  \beta < 2$ at low 
temperature $(T <150 K)$  \cite{jang}. A recent review of thermal properties of graphene and nanostructured carbon materials can be found in \cite{nature}. 
 
Our model is presented to describe the heat conduction of any kind of  solid state without additional restrictions, therefore 
room or even higher temperature can be considered with large negative $\omega$ exponents.   

Even from these examples we can see that it has a need to investigate the general heat conduction problem, where 
the coefficients have general power law dependence.  

We look for the solutions of (\ref{eqn0},\ref{eqn1}) in the most general self-similar form
\begin{equation}
T=t^{-\alpha}f(\eta),  
\hspace{1cm} q=t^{-\delta}g(\eta).  
\label{eqn2}
\end{equation} 
 For a better transparency in the following we introduce a new variable $\eta = \frac{r}{t^\beta}$, 
where $\alpha,\beta,\delta $ are all real numbers. 
 
The similarity exponents $\alpha, \delta$ and $\beta$ are of primary physical importance since 
$\alpha, \delta$  represents the rate of decay of the magnitude T or q, 
while $\beta$  is the rate of spread 
(or contraction if  $\beta<0$ ) of the space distribution as time goes on. 
Self-similar solutions exclude the existence of any single time scale in the investigated system.

We substitute (\ref{eqn2}) into (\ref{eqn0}) and (\ref{eqn1}). It can be checked  that 
\eq
\alpha = \frac{1}{\omega+1}, \hspace*{3mm} \beta = \frac{1}{2(\omega+1)}, \hspace*{3mm} \delta =\frac{2\omega+3}{2(\omega+1)} , \hspace*{3mm} 
\epsilon = \omega+1. 
\eqe 
Then we can obtain the shape functions f and g the following ordinary differential equation (ODE) system  
\begin{eqnarray} 
\delta g + \beta \eta g' &=&  g f^{\omega+1} + f^{2\omega+1}f',
  \label{Eqn3} \\ 
(\eta g)' &=&  \beta(\eta^2 f)' 
\label{Eqn4}
\end{eqnarray}
where prime means derivation with respect to $\eta$. 
 
The first lucky moment is that (\ref{Eqn4}) relates f and g in a simple way 
\eq 
g =  \beta \eta f 
\label{copl}  
\eqe 
if the  $\alpha = 2\beta $ universality relation is fulfilled.


Note, that we can immediately read how the self-similar solutions of the temperature distribution T and the heat 
flux q depend on $\omega$
\eq
T  = t^{\frac{-1}{\omega+1}} f \left( \frac{r}{t^{ \frac{1}{2(\omega+1)}  }} \right),  \hspace*{8mm}  q = 
 t^{\frac{2 \omega +3}{2(\omega+1)}} g \left( \frac{r}{t^{ \frac{1}{2(\omega+1)}  }} \right). 
\eqe 
The parameter dependence of the complete heat conduction coefficient and relaxation time can be expressed via $\omega$ 
as well
\eq
\kappa   = \kappa_0  t^{\frac{-\omega}{\omega+1}} f^{\omega} \left( \frac{r}{t^{ \frac{1}{2(\omega+1)}  }} \right),     \hspace{8mm} \tau =  \kappa_0  t^{-1}  f^{\omega +1} \left(  \frac{r}{t^{ \frac{1}{2(\omega+1)}}  } \right).
\eqe 
Recall that  $\omega > -1$. 
These are already very informative and useful relations to investigate the global properties of the solutions, 
note that such kind of analysis are available for large number of complex mechanical and flow problems \cite{sed}. 

Substituting these relations back to Eq. (\ref{Eqn3}) after some algebra 
we arrive at the following non-linear first-order ODE   
\eq
\frac{df}{d\eta^2} \left(\beta^2 \eta^2 - f^{2\omega+1} \right) = \frac{\beta f}{2} [f^{\omega+1}-(2\beta+1)].
 \label{direct} 
\eqe 
Put $y = \eta^2 $ and $ x =f$. With this notation eq. (16) becomes linear for $ y(x)$ (this is the second lucky moment
of investigation): 

\eq
\frac{dy}{dx} =  \frac{y(x)-4(\omega+1)^2x^{2\omega+1} }{x[(\omega+1)x^{\omega+1}-\omega-2]}.
\label{invers}
\eqe

 Plainly, $f \equiv 0$ is a solution to eq. (16). If $y(x) $ the solution of eq. (17) is strictly monotonic then 
so is the inverse function $f=x$ and no discontinuity. However if $y(x)$ is not monotonic on some interval $(x_1,x_2)$ 
and has a turning point at $x_0 \epsilon (x_1,x_2)$ then the inverse $(f=x)$ has sense on $[0,y(x)]$ only. 
One sets $f=0$ for $y> y(x_0)$ and the discontinuity a $y(x_0)$ is apparent. 
 The analytical investigation of the linear equation (17) is, in general easier than of eq (16). 
In some cases (for some $\omega$s) one can have more explicit or almost explicit solutions.  

There are two examples: \\ 
The first case is for $\omega =0,  ( \alpha = 1, \beta =1/2, \delta = 3/2, \epsilon = 1 )$.  \\ 
 This example was studied 
by \cite{choi} in some details. The corresponding ODE (17) reads $y' = (y-4x)/x(x-2) $ which has a solution 
 $y = 8 + [(x-2)/x]^{1/2} [c_1 - 8 ln(\sqrt{x}+\sqrt{x-2} )] $ where   $c_1$ is a constant.  

 It is clear that must be $x\ge 2$ and $y(x)$ is monotonic for $x>2$ until $x_0$ where $y=0$. This means 
that $x(y)$ exists and monotonic on some interval $[0,y_0], x(y_0) =2;$ 
for $y \ge y_0$ we have $x(y) =0$ so the discontinuity. For a better understanding 
Figure 1a presents the graph of solution of Eq. (17) through the point (3,0.5). The inverse of this function for $x>2$ 
is shown on Fig. 1b (the nonzero part). The solid line is a solution through the $f(0)=10.8 $ point.  Figure 2 presents the shock-wave propagation of the temperature distribution  $T(r,t)$ for $\omega = 0 $. 

The second case is for $\omega = -1/2,  (\alpha = 2, \beta =1, \delta = 2, \epsilon = 1/2)$.  \\ 
Now Eq. (17) takes the form of 
$ \frac{dy}{dx} =  2(y-1)/[x(\sqrt{x}-3)]$.
It can be checked that $y = c_2 x^{-2/3}(x^{1/2}-3)^{4/3}$
is a solution for any $c_2 > 0.$ Take $c_2 = 1$. The function $y(x)$ is monotonic 
on $ (0,9),  y(9) = 0  $. Returning to original variables we have 
$f = 9/[(\eta^2)^{3/4}+1]^2$  (which is plainly less than 9!) 
According to eq. (14) temperature and heat flux distributions are 
\eq
 T =    \frac{9t}{(r^{3/2} +t^{3/2})^2  }, \hspace*{2cm}  
q =   \frac{9r}{(r^{3/2} +t^{3/2})^2  }. 
\label{distr}
\eqe
These solutions are \underline{not} discontinuous. 
Analytical and numerical calculus suggest that $\omega = -1/2$ is a critical exponent: for $-1 < \omega \le -1/2 $ 
the solutions are continuous, for $\omega  > -1/2$ the shocks always appear.  

{\it{In summary}} \\
We presented a hyperbolic model for heat conduction in solids where the relaxation time and heat conduction coefficient 
is a power law function of time. There are basically two different regimes available for different power laws. 
For $\-1 < \omega \le -1/2 $ the solutions are continuous for all positive time and radial coordinate, 
for  $    \omega > -1/2 $ the solutions are only continuous on a finite and closed $  [0:\eta_0]$  interval and have a finite jump at the the endpoint $\eta_0$. As physical interpretation numerous materials and solid state systems mentioned with 
temperature dependent heat conduction coefficients. \\ 
The paper is dedicated to Annabella Barna who was born on 20th of December 2011. 
 
\begin{figure}
    \label{fig:subfigures}
    \begin{center}
{{\scalebox{0.45}{\rotatebox{0}{\includegraphics{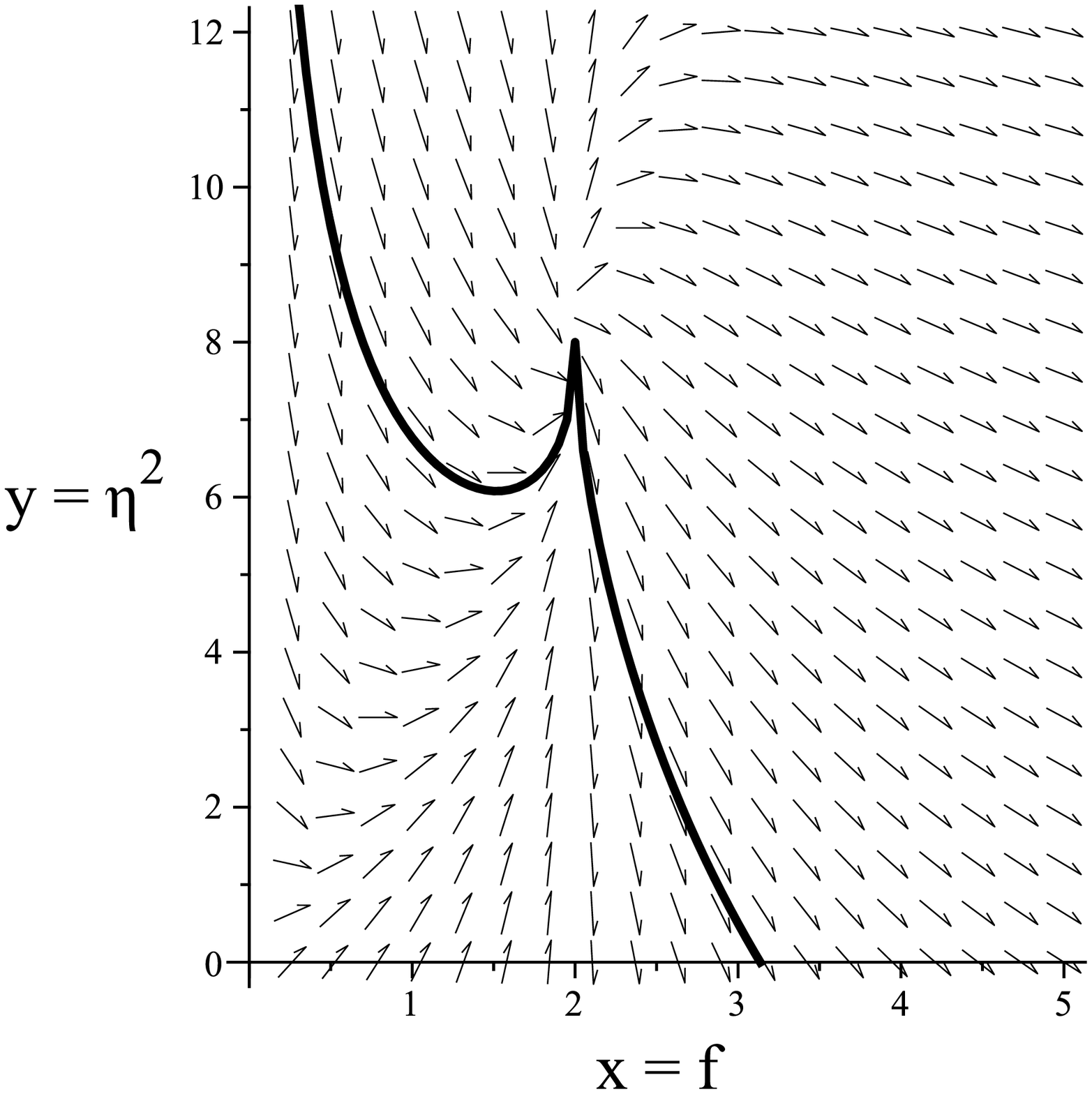}}}} }
{{\scalebox{0.45}{\rotatebox{0}{\includegraphics{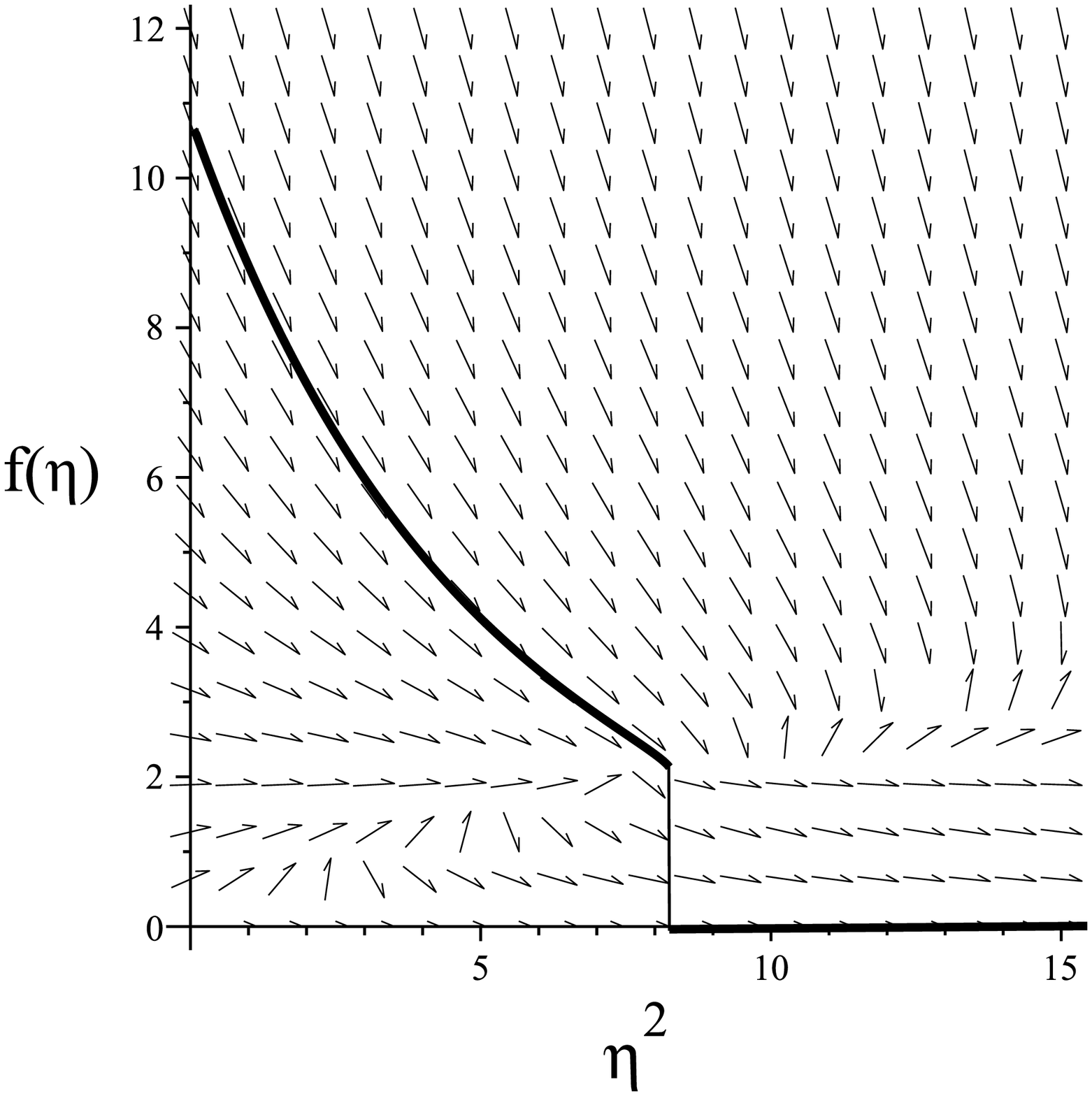}}}} }
a)   \hspace*{3cm} b)
    \end{center}
    \caption{The direction field of   a)   Eq. (17) for  $\omega = 0$ and b) Eq. (16) for $\omega =0$
 The solid line presents numerical solutions for a)  $y(3)=0.5$ and for b)   $f(0)=10.8$. 
     }%
\end{figure}
\begin{figure}
    \label{fig:subfigures}
    \begin{center}
{\rotatebox{-90}{\scalebox{0.77}{\includegraphics{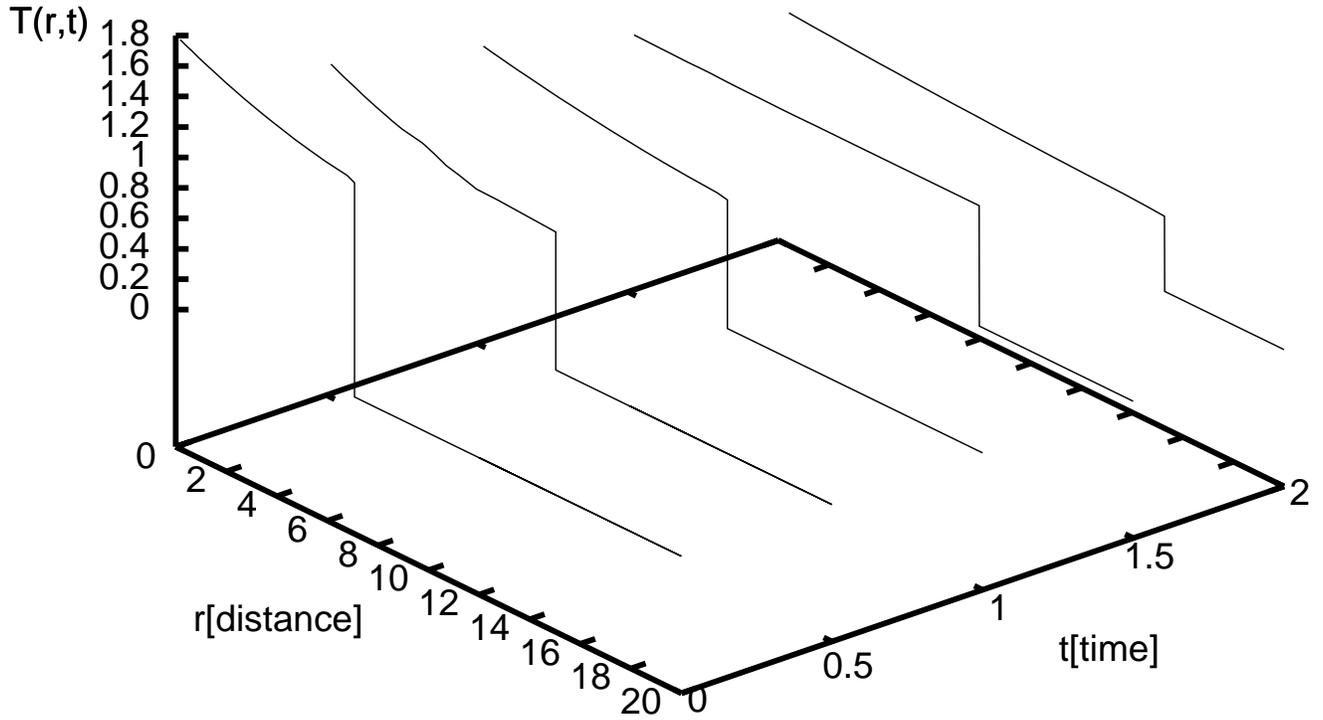}}}}  
    \end{center}
    \caption{ The shock-wave propagation of the temperature distribution of $T(r,t)$ for $\omega =0$}
\end{figure}

                                                                  
\end{document}